\documentclass[usenatbib]{mn2e}
\usepackage{graphicx,times}
\usepackage{bm}
\graphicspath{{./fig/}{./png/}}
\newcommand{\EQ}{\begin{equation}}
\newcommand{\EN}{\end{equation}}
\newcommand{\EQA}{\begin{eqnarray}}
\newcommand{\ENA}{\end{eqnarray}}
\newcommand{\eq}[1]{(\ref{#1})}

\newcommand{\Eq}[1]{equation~(\ref{#1})}
\newcommand{\Eqs}[2]{equations~(\ref{#1}) and~(\ref{#2})}
\newcommand{\Eqss}[2]{equations~(\ref{#1})--(\ref{#2})}

\newcommand{\Sec}[1]{Sect.~\ref{#1}}

\newcommand{\Fig}[1]{Fig.~\ref{#1}}

\newcommand{\Tab}[1]{Table~\ref{#1}}

\newcommand{\bra}[1]{\langle #1\rangle}

{}
{}
{}

{}
{}
\newcommand{\meanEMF}{\overline{\mbox{\boldmath ${\cal E}$}}{}}{}
{}
{}
{}
{}
{}
\newcommand{\meanBB}{\overline{\mbox{\boldmath $B$}}{}}{}
{}
{}
{}
{}
{}
{}
{}
\newcommand{\meanJJ}{\overline{\mbox{\boldmath $J$}}{}}{}
\newcommand{\meanUU}{\overline{\mbox{\boldmath $U$}}{}}{}
{}
{}
{}

\newcommand{\meanB}{\overline{B}}

{}
\newcommand{\hatBB}{\hat{\bm{B}}}
{}

{}
{}

\newcommand{\hatB}{\hat{B}}

\newcommand{\zzz}{\hat{\mbox{\boldmath $z$}} {}}

\newcommand{\nullvector}{{\bf0}}

\newcommand{\kk}{\bm{k}}

\newcommand{\uu}{\mbox{\boldmath $u$} {}}
\newcommand{\UU}{\mbox{\boldmath $U$} {}}

\newcommand{\aaa}{\mbox{\boldmath $a$} {}}
\newcommand{\bb}{\mbox{\boldmath $b$} {}}
\newcommand{\BB}{\mbox{\boldmath $B$} {}}

\newcommand{\JJ}{\mbox{\boldmath $J$} {}}

\newcommand{\AAA}{\mbox{\boldmath $A$} {}}

\newcommand{\ff}{\mbox{\boldmath $f$} {}}

\newcommand{\FF}{\mbox{\boldmath $F$} {}}

\newcommand{\nab}{\mbox{\boldmath $\nabla$} {}}

\newcommand{\SSSS}{\mbox{\boldmath ${\sf S}$} {}}

\newcommand{\EMF}{\mbox{\boldmath ${\cal E}$} {}}

\newcommand{\DD}{{\rm D} {}}

\newcommand{\dd}{{\rm d} {}}
\newcommand{\const}{{\rm const}  {}}

\def\ga{\mathrel{\mathchoice {\vcenter{\offinterlineskip\halign{\hfil
$\displaystyle##$\hfil\cr>\cr\sim\cr}}}
{\vcenter{\offinterlineskip\halign{\hfil$\textstyle##$\hfil\cr>\cr\sim\cr}}}
{\vcenter{\offinterlineskip\halign{\hfil$\scriptstyle##$\hfil\cr>\cr\sim\cr}}}
{\vcenter{\offinterlineskip\halign{\hfil$\scriptscriptstyle##$\hfil\cr>\cr\sim\cr}}}}}
\def\Ma{\mbox{\rm Ma}}

\def\Pm{P_{\rm m}}
\def\Rm{R_{\rm m}}

\def\cs{c_{\rm s}}

\def\kf{k_{\rm f}}

\def\urms{u_{\rm rms}}

\def\nut{\nu_{\rm t}}
\def\etat{\eta_{\rm t}}

\def\Beq{B_{\rm eq}}

\def\half{{\textstyle{1\over2}}}

\def\onethird{{\textstyle{1\over3}}}

\newcommand{\yapj}[3]{ #1, {ApJ,} {#2}, #3}

\newcommand{\yan}[3]{ #1, {Astron.\ Nachr.,} {#2}, #3}

\newcommand{\yana}[3]{ #1, {A\&A,} {#2}, #3}

\newcommand{\ygafd}[3]{ #1, {Geophys.\ Astrophys.\ Fluid Dyn.,} {#2}, #3}

\newcommand{\yjetp}[3]{ #1, {Sov.\ Phys.\ JETP,} {#2}, #3}

\newcommand{\yprl}[3]{ #1, {Phys.\ Rev.\ Lett.,} {#2}, #3}

\newcommand{\ymn}[3]{ #1, {MNRAS,} {#2}, #3}

\newcommand{\ypre}[3]{ #1, {Phys.\ Rev.\ E,} {#2}, #3}

\newcommand{\yjour}[4]{ #1, {#2}, {#3}, #4}

\newcommand{\ybook}[3]{ #1, {#2} (#3)}

\title[The $\alpha$ effect with imposed magnetic fields]
{The $\alpha$ effect with imposed and dynamo-generated magnetic fields}
\author[A.\ Hubbard et al.]{A.\ Hubbard$^{1}$, F.\ Del Sordo$^{1,2}$,
P.\ J.\ K\"apyl\"a$^{3}$ and A.\ Brandenburg$^{1,2}$\\
$^1$NORDITA, AlbaNova University Center, Roslagstullsbacken 23,
SE-10691 Stockholm, Sweden\\
$^2$Department of Astronomy, AlbaNova University Center,
Stockholm University, SE-10691 Stockholm, Sweden\\
$^3$Observatory, T\"ahtitorninm\"aki (PO Box 14),
FI-00014 University of Helsinki, Finland
}
\date{Accepted 2009 May 12. Received 2009 May 12; in original form 2009 April 20}
\begin{document}

\maketitle

\begin{abstract}
Estimates for the nonlinear $\alpha$ effect in helical turbulence
with an applied magnetic field are presented using two different
approaches: the imposed-field method where the electromotive force owing
to the applied field is used, and the test-field method where separate
evolution equations are solved for a set of different test fields.
Both approaches agree for stronger fields, but there are apparent
discrepancies for weaker fields that can be explained by the influence
of dynamo-generated magnetic fields on the scale of the domain
that are referred to as meso-scale magnetic fields.
Examples are discussed where these meso-scale fields can lead to 
both drastically overestimated and underestimated values of $\alpha$
compared with the kinematic case.
It is demonstrated that the kinematic value can be recovered by
resetting the fluctuating magnetic field to zero in regular time intervals.
It is concluded that this is the preferred technique both for the
imposed-field and the test-field methods.
\end{abstract}
\label{firstpage}

\section{Introduction}

The $\alpha$ effect is commonly used to describe the evolution of the
large-scale magnetic field in hydromagnetic dynamos \citep{Mof78,Par79,KR80}.
However, the $\alpha$ effect is not the only known mechanism for
explaining the generation of large-scale magnetic fields.
Two more effects have been discussed in cases when there is shear
in the system:
the incoherent alpha--shear dynamo \citep{VB97,Sok97,Sil00,Pro07}
and the shear--current effect \citep{RK03,RK04}.
In order to provide some understanding of the magnetic field generation
in astrophysical bodies such as the Sun or the Galaxy, or at least in
numerical simulations of these systems, it is of interest to be able
to identify the underlying mechanism.

Astrophysical dynamos are usually confined to finite
domains harboring turbulent fluid motion.
Both the Sun and the Galaxy are gravitationally stratified and rotating,
which makes the turbulence non-mirror symmetric, thus leading to an $\alpha$ effect.
In addition, the rotation is nonuniform, which leads to a strong
amplification of the magnetic field in the toroidal direction,
as well as other effects such as those mentioned above.
Instead of simulating such systems with all their ingredients, it
is useful to simplify the setup by restricting oneself to
Cartesian domains that can be thought to represent a part of
the full domain.
At low magnetic Reynolds numbers, i.e.\ when the effects of
induction are comparable to those of magnetic diffusion,
the $\alpha$ effect can clearly be identified in simulations
of convection in Cartesian domains; see \cite{Betal90}.
Here, $\alpha$ has been determined by applying a
uniform magnetic field across the simulation domain
and measuring the resulting electromotive force.
This method is referred to as the imposed-field method.
However, in subsequent years simulations at larger magnetic
Reynolds numbers have revealed problems in that the resulting
$\alpha$ becomes smaller and strongly fluctuating in time.
This was first found in simulations where the turbulence is caused
by an externally imposed body force \citep{CH96}, but
it was later also found for convection \citep{CH06}.
This suggested that the mean-field approach may be
seriously flawed \citep{CH09}.

Meanwhile, there have been a number of simulations of convection
where large-scale magnetic fields are being generated.
Such systems include not only simulations in spherical shells
\citep{Browning,Brown}, but also in Cartesian
domains \citep{Ketal08,Ketal09a,HP09}.
However, the absence of a significant $\alpha$ effect in some of
these simulations led \cite{HP09} to the conclusion
that such magnetic fields can only be explained by other mechanisms
such as the incoherent alpha--shear dynamo or the shear--current effect.
Such an explanation seems to be in conflict with earlier claims of a finite
$\alpha$ effect as determined by the test-field method of
\citep{Sch05,Sch07}, and in particular with recent results
for convection \citep{Ketal09b}.
The purpose of the present paper is therefore to discuss possible
reasons for conflicting results that are based on different methods.
The idea is to compare measurements of the $\alpha$ effect
using both the imposed-field method and the test-field method.
We consider here the case of helically forced turbulence
in a triply-periodic domain.
This case is believed to be well understood.
We expect $\alpha$ to be catastrophically quenched, i.e.\ $\alpha$
is suppressed for field strengths exceeding the \cite{Zel57} value
of $\Rm^{-1/2}\Beq$, where $\Beq$ is the equipartition field strength
where kinetic and magnetic energy densities are comparable.
The importance of the Zeldovich field strength was emphasized
by \cite{GD94} in connection with catastrophic quenching resulting
from magnetic helicity conservation.

In this paper we focus on the case of moderate values of $\Rm$ of
around 30.
This is small by comparison with astrophysical applications, but it
is large compared with the critical value for dynamo action in
fully helical turbulence \citep{B01}, which
occurs for $\Rm\ga1$ in our definition of $\Rm$ based on the wavenumber
of the scale of the energy-carrying eddies, i.e.\ the forcing wavenumber.
In addition, we only consider cases with a magnetic Prandtl number of unity.
However, this should not worry us too much, because we know that the
large-scale dynamo works independently of the value of the
magnetic Prandtl number \citep{Min07,B09}.

\section{Helical turbulence and $\alpha$ effect}

\subsection{Forced turbulence simulations}

Throughout this paper we consider hydromagnetic turbulence in the presence
of a mean magnetic field $\BB_0$ using triply-periodic boundary conditions.
The total magnetic field is written as $\BB_0+\nab\times\AAA$, where
$\AAA$ is the magnetic vector potential.
We employ an isothermal equation of state where the pressure is proportional
to the density, $p=\rho\cs^2$, with $\cs$ being the isothermal sound speed.
The governing evolution equations for logarithmic density $\ln\rho$,
velocity $\UU$, together with $\AAA$, are given by
\EQ
{\DD\ln\rho\over\DD t}=-\nab\cdot\UU,
\label{dlnrhodt}
\EN
\EQ
{\DD\UU\over\DD t}=\JJ\times(\BB_0+\BB)/\rho
+\ff+\FF_{\rm visc}-\cs^2\nab\ln\rho,
\label{dUdt}
\EN
\EQ
{\partial\AAA\over\partial t}=\UU\times(\BB_0+\BB)+\eta\nabla^2\AAA,
\label{dAdt}
\EN
where $\BB_0+\BB$ is the total magnetic field, but since $\BB_0=\const$
it does not enter in the mean current density, which is given by
$\JJ=\nab\times\BB/\mu_0$, where $\mu_0$ is the vacuum permeability.
Furthermore, $\DD/\DD t=\partial/\partial t+\UU\cdot\nab$ is the advective
derivative, $\FF_{\rm visc}=\rho^{-1}\nab\cdot2\rho\nu\SSSS$ is the viscous
force, $\nu$ is the kinematic viscosity,
${\sf S}_{ij}=\half(U_{i,j}+U_{j,i})-\onethird\delta_{ij}\nab\cdot\UU$
is the traceless rate of strain tensor, and $\ff$ is a random forcing
function consisting of plane transversal waves with random wavevectors
$\kk$ such that $|\kk|$ lies in a band around a given forcing wavenumber
$k_{\rm f}$.
The vector $\kk$ changes randomly from one timestep to the next.
This method is described for example in \cite{HBD04}.
The forcing amplitude is chosen so that the Mach number $\Ma=\urms/\cs$
is about 0.1.

We consider a domain of size $L_x\times L_y\times L_z$.
We use $L_x=L_y=L_z=2\pi/k_1$ in all cases.
Our model is characterized by the choice of magnetic Reynolds
and Prandtl numbers, defined here via
\EQ
\Rm=\urms/\eta\kf,\quad
\Pm=\nu/\eta.
\EN
We start the simulations with zero initial magnetic field,
so the field is entirely produced by the imposed field.
The value of the magnetic field will be expressed in units
of the equipartition value
\EQ
B_{\rm eq}=\bra{\mu_0\rho\uu^2}^{1/2}.
\EN
We consider values of $B_0/\Beq$ from 0.06 to 20 along with a
magnetic Reynolds number of about 26, adequate to support dynamo action.

\subsection{$\alpha$ from the imposed-field method}

The present simulations allow us to determine directly the $\alpha$ effect
under the assumption that the relevant mean field is given by volume averages,
denoted here by angular brackets.
Given that the magnetic field is written as $\BB=\nab\times\AAA$ where $\AAA$ is
also triply periodic, we have $\bra{\BB}=\nullvector$.
We can determine the volume-averaged electromotive force,
\EQ
\bra{\EMF}=\bra{\EMF}(t)\equiv\bra{\uu\times\bb},
\EN
where $\uu=\UU-\bra{\UU}$ and $\bb=\BB$ are the
fluctuating components of velocity and magnetic field, and
$\bra{\BB}=\bra{\nab\times\AAA}=\nullvector$.

For mean fields defined as volume averages, and because
of periodic boundary conditions, we have $\bra{\JJ}={\bf0}$.
Under isotropic conditions there is therefore
only the $\alpha$ effect connecting $\bra{\EMF}$
with $\BB_0$ via $\bra{\EMF}=\alpha_{\rm imp}\BB_{0}$, so
\EQ
\alpha_{\rm imp}=\bra{\EMF}\cdot\BB_{0}/B_0^2.
\EN
In all cases reported below we assume $\BB_0=(B_0,0,0)$.
Note that $\nab\times\bra{\EMF}=\nullvector$ and therefore
our time-constant imposed field is self-consistent.

\subsection{$\alpha$ from the test-field method}
\label{testfield}

A favored method of determining the full $\alpha_{ij}$ tensor is
by using the test-field method \citep{Sch05,Sch07}, where one solves, in addition to
\Eqss{dlnrhodt}{dAdt}, a set of equations.
In the special case of volume averages this set of equations simplifies to
\EQ
{\partial\aaa^{q}\over\partial t}\!=\!\meanUU\times\bb^{q}+\uu\times(\BB_0+\meanBB^{q})
+\uu\times\bb^{q}-\overline{\uu\times\bb^{q}}+\eta\nabla^2\aaa^{q},
\label{nonSOCA}
\EN
where $\bb^q=\nab\times\aaa^q$ with $q=1$ or 2 denotes the response to
each of the two test fields $\meanBB^{q}$.
Throughout this paper, overbars denote planar averages.
Later we consider arbitrary planar averages and denote their normals
by superscripts, but here we restrict ourselves to $xy$ averages.
We use two different constant test fields,
\EQ
\meanBB^1=({\cal B},0,0),\quad
\meanBB^2=(0,{\cal B},0),
\EN
where ${\cal B}=\const$ is the magnitude of the test field,
but its actual value is of no direct significance, because the ${\cal B}$
factor cancels in the calculation of $\alpha$.

However, given that the test-field equations are linear in $\bb^q$,
this field can grow exponentially due to dynamo action.
When $|\bb^q|$ becomes larger than about 20 times the value of ${\cal B}$,
the determination of $\alpha$ becomes increasingly inaccurate,
so it is advisable to reset $\bb^q$ to zero in regular intervals
\citep{Sur_etal08}.
We calculate the corresponding values of the electromotive force
$\bra{\EMF}^q=\bra{\uu\times\bb^q}$ to determine the components
\EQ
\alpha_{iq}=\bra{\EMF}_i^q/{\cal B}.
\EN
This corresponds to the special case $k=0$ when considering sinusoidal
and cosinusoidal test functions described elsewhere \citep{BRS08}.

Even though the test-field equations themselves are linear, the
flow field is affected by the actual magnetic field (which is different
from the test field), so the resulting $\alpha$ tensor is being
affected (``quenched'') by the magnetic field.
This was successfully demonstrated in \cite{BRRS08},
where $\alpha_{ij}$ takes the form
\EQ
\alpha_{ij}=\alpha_1\delta_{ij}+\alpha_2\hatB_i\hatB_j.
\label{alpij}
\EN
Here $\hatBB=\meanBB/|\meanBB|$ is the unit vector of the
relevant mean magnetic field.
In the induction equation the $\alpha$ effect occurs only in the combination
\EQ
\alpha_{ij}\meanB_j=(\alpha_1+\alpha_2)\meanB_i,
\EN
and this is also what is determined by the imposed-field method,
but it is different from the mean values of the components of the
$\alpha_{ij}$ tensor.
On the other hand, in the case of a passive vector field it is the mean
components of $\alpha_{ij}$ rather than the components of
$\alpha_{ij}\meanB_j$ that are of immediate importance \citep{TB08}.

\subsection{$\alpha$ in the presence of meso-scale fields}

The {\it relevant} mean field may not just be the imposed field
with wavenumber $k=0$, but it may well be a field with wavenumber $k=k_1$.
Such a field would vanish under volume averaging, but it would still
produce finite values of $\bra{\hatB_i\hatB_j}$.
For the diagonal components of $\bra{\alpha_{ij}}$ we can write
\EQ
\bra{\alpha_{xx}}=\alpha_1+\epsilon_x\alpha_2,\quad
\bra{\alpha_{yy}}=\alpha_1+\epsilon_y\alpha_2,
\EN
where the factors
\EQ
\epsilon_x=\bra{\hatB_x^2},\quad\mbox{and}\quad
\epsilon_y=\bra{\hatB_y^2},
\EN
quantify the weight of the $\alpha_2$ term.
For a purely uniform field pointing in the $x$ direction we have
$\epsilon_x=1$ and $\epsilon_y=0$, while for a Beltrami field of the form
$\meanBB=(\cos kz, \sin kz, 0)$ we have $\epsilon_x=\epsilon_y=1/2$.

In practice we will have a mixture between the imposed field
(below sometimes referred to as large-scale field) and a dynamo-generated
magnetic field with typical wavenumber $k=k_1$ (below sometimes referred
to as meso-scale magnetic field).
The solution to the test-field equations, $\bb^q$, can also develop
meso-scale fields with wavevectors in the $x$ or $y$ directions, but
not in the $z$ direction, because that component is removed by the
term $\overline{\uu\times\bb^{q}}$ in equation~(\ref{nonSOCA}).
\Tab{VariousFields} highlights the difference between imposed,
meso-scale, and test fields.
We denote the ratio of the strengths of imposed and meso-scale fields
as $\beta=B_0/B_1$
and distinguish three (and later four) different cases, depending on the direction
of the wavevector of the Beltrami field.

\begin{table}\caption{
Overview of the different types of fields and their meaning.
}\vspace{12pt}\centerline{\begin{tabular}{lcccc}
field & symbol & magn & induct. eqn & test-field eqn \\
\hline
imposed field    & $\BB_0$     & $B_0$      & yes  &  yes  \\
meso-scale field & $\meanBB$   & $B_1$      & yes  &  ---  \\
test field       & $\meanBB^q$ & ${\cal B}$ & ---  &  yes  \\
test field response & $\bb^q$   &            & ---  &  yes  \\
\label{VariousFields}\end{tabular}}\end{table}

The first case is referred to as the X branch, because the wavevector
of the Beltrami field points in the $x$ direction.
To calculate $\epsilon_x$ there is, in addition to the imposed field
$B_0$, a Beltrami field $B_1(0,\cos kx,\sin kx)$, which does not have a
component in the $x$ direction.
Thus, $B_x=B_0$, and since $\BB=(B_0,B_1\cos kx,B_1\sin kx)$, we have
$\BB^2=B_0^2+B_1^2$, so $\epsilon_x=\hatB_x^2=B_0^2/(B_0^2+B_1^2)$, or
$\epsilon_x=\beta^2/(1+\beta^2)$.
Likewise, with $B_y=B_1\cos kx$ we find for the volume average or,
in this case, the $x$ average $\bra{B_y^2}=B_1^2/2$, so
$\epsilon_y=1/[2(1+\beta^2)]$.

The next case is referred to as the Y branch, because the wavevector
of the Beltrami field points in the $y$ direction.
Thus, we have $\BB=(B_0+B_1\sin ky,0,B_1\cos ky)$, so
$\BB^2=B_0^2+2B_0 B_1\sin ky+B_1^2$.
This is no longer independent of position, so the volume average or, in this case,
the $y$ average has to be obtained by integration.
Thus, we write $\epsilon_x=I_1(\beta)$ where we have defined
\begin{eqnarray}
\nonumber
I_1(\beta)=\int_0^{2\pi}\!\!
{(\beta+\sin\theta)^2\over\beta^2+2\beta\sin\theta+1}\,\dd\theta
=\left\{\!\!
 \begin{array}{lc}
1/2 & \beta^2 \leq1, \\ 
1-1/2\beta^2 & \beta^2>1,
\end{array} \right.
\end{eqnarray}
where $\theta=ky$ has been introduced as dummy variable.
Since $B_y=0$ in this case, we have $\epsilon_y=0$.

\begin{figure}\begin{center}
\includegraphics[width=\columnwidth]{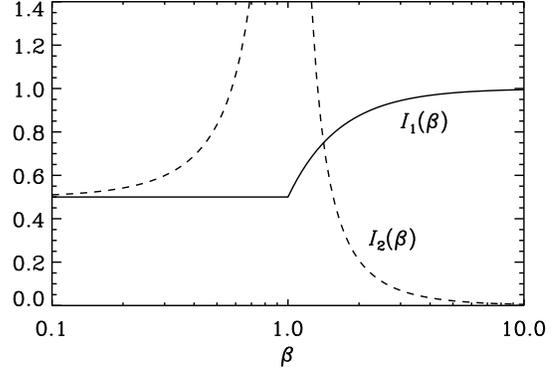}
\end{center}\caption[]{
Plot of the integrals $I_1(\beta)$ and $I_2(\beta)$.

}\label{paver}\end{figure}

Finally for the Z branch, where the wavevector
of the Beltrami field points in the $z$ direction, we have
$\BB=(B_0+B_1\cos kz,B_1\sin kz,0)$, we find
$\epsilon_x=I_1(\beta)$ and $\epsilon_y=I_2(\beta)$ with
\begin{eqnarray} 
\nonumber
I_2(\beta)=\int_{0}^{2\pi}\!
{\cos^2\theta\over\beta^2+2\beta\cos\theta +1}\,{\dd\theta\over2\pi}
=\left\{\!\!
 \begin{array}{lc}
I_0(\beta) & \beta^2<1, \\ 
I_0(\beta)/\beta^2 & \beta^2>1,
\end{array} \right.
\end{eqnarray}
where $I_0(\beta)=(1+\beta^2)/[2(1-\beta^2)]$ and
$\theta=kz$ has been used as a dummy variable.
A graphical representation of the integrals is given in \Fig{paver} and a
summary of the expressions for $\epsilon_x(\beta)$ and $\epsilon_y(\beta)$
as well as $\epsilon_x(0)$ and $\epsilon_y(0)$ for the X, Y, and Z
branches is given in \Tab{TSum}.
The singularity in $I_0(\beta)$ could potentially affect $\alpha_{yy}$.
However, the results shown below show that, at least for stronger fields,
$\alpha_2$ goes to zero near the singularity of $I_0(\beta)$ such that
$\alpha_{yy}$ remains finite.

\begin{figure*}\begin{center}
\includegraphics[width=\textwidth]{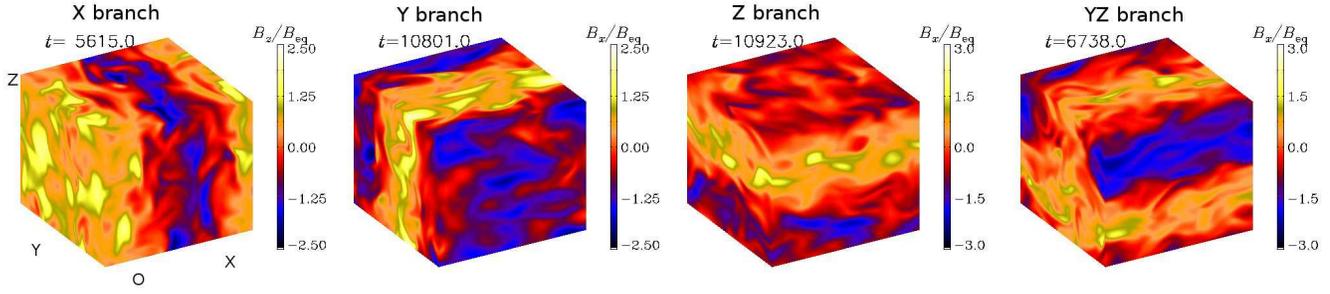}
\end{center}\caption[]{
Visualization of $B_z$ on the periphery of the computational domain
for the X branch and $B_x$ for the Y, Z, and YZ branches.
The coordinate directions are indicated on the first panel.
}\label{XYZ}\end{figure*}

\begin{table}\caption{
Summary of the expressions for $\epsilon_x(\beta)$ and $\epsilon_y(\beta)$
as well as $\epsilon_x(0)$ and $\epsilon_y(0)$ for the X, Y, and Z branches.
}\vspace{12pt}\centerline{\begin{tabular}{ccccc}
Branch & $\epsilon_x(\beta)$ & $\epsilon_y(\beta)$
       & $\epsilon_x(0)$ & $\epsilon_y(0)$ \\
\hline
X & $\beta^2/(1+\beta^2)$ & $1/[2(1+\beta^2)]$ & 0 & 1/2 \\
Y & $I_1(\beta)$ & $0$ & 1/2 & 0 \\
Z & $I_1(\beta)$ & $I_2(\beta)$ & 1/2 & 1/2 \\
\label{TSum}\end{tabular}}\end{table}

\section{Results}

We have performed simulations for values of $B_0$ in
the range $0.06\leq \Rm^{1/2}B_0/\Beq\leq20$ for $\Rm\approx26$ and $\Pm=1$.
In all cases we use $\kf/k_1=3$, which is big enough to allow
a meso-scale magnetic field of wavenumber $k_1$ to develop
within the domain; see \Fig{XYZ}.
We did not initially anticipate the importance of the meso-scale fields.
Different runs were found to exhibit rather different behavior which
turned out to be related to their random positioning on different branches.
We used the existing results from different branches as initial conditions
for neighboring values of $B_0$.

In this paper, error bars are estimated from the averages obtained from
any of three equally long subsections of the full time series.
The error bars are comparable with the typical scatter of the data points,
but they are not shown because they would make the figure harder to read.
Note that the results in this section consider saturated fields.
The opposite case will be considered in \Sec{Resetting}.

\subsection{Different branches}

\begin{figure}\begin{center}
\includegraphics[width=\columnwidth]{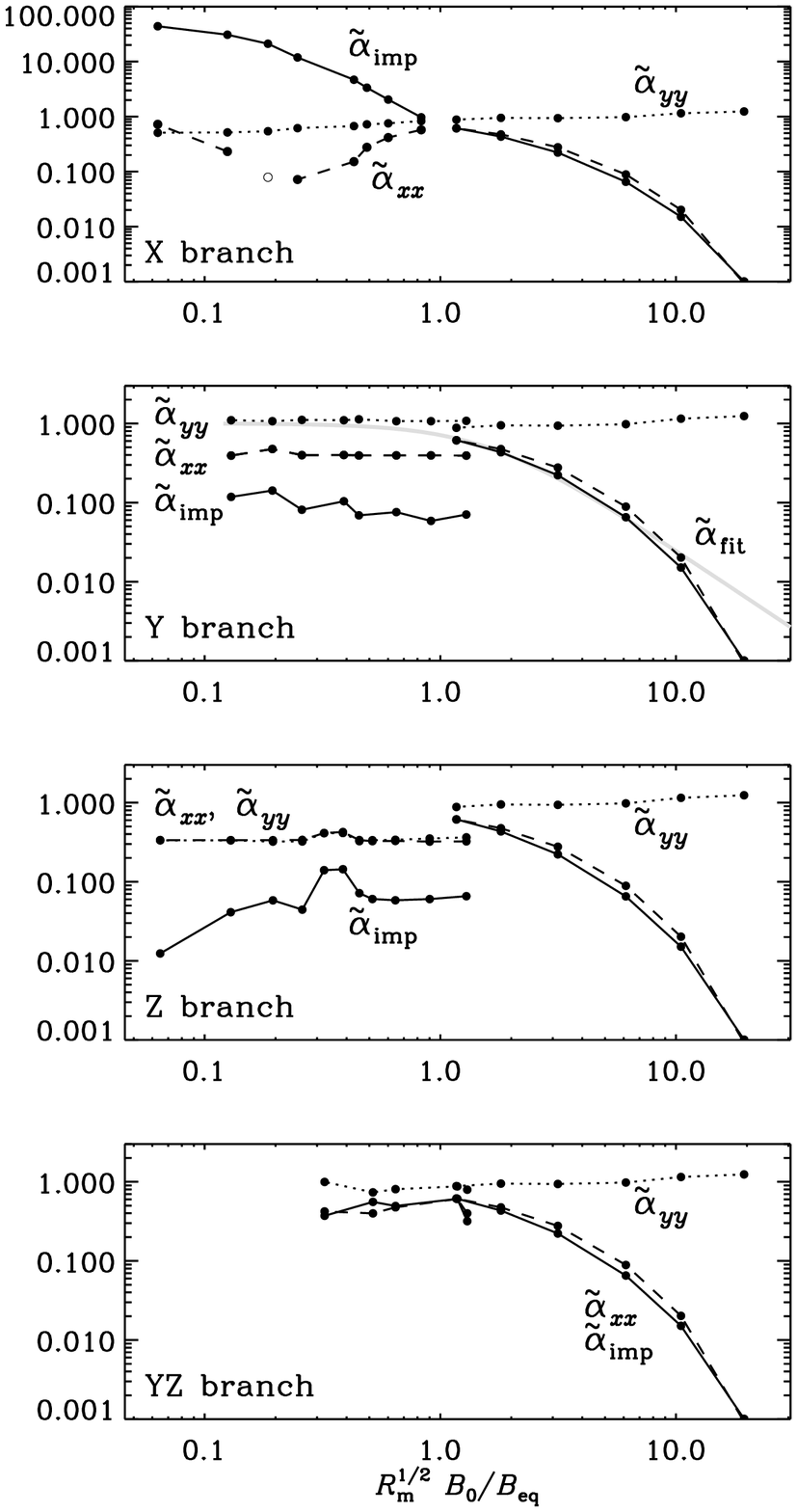}
\end{center}\caption[]{
Volume-averaged values of $\alpha_{\rm imp}$, $\alpha_{xx}$,
and $\alpha_{yy}$.
A tilde indicates that the values are normalized by $\alpha_0$,
i.e.\ $\tilde{\alpha}_{\rm imp}=\alpha_{\rm  imp}/\alpha_0$ (solid line),
$\tilde{\alpha}_{xx}=\bra{\alpha_{xx}}/\alpha_0$ (dashed line),
$\tilde{\alpha}_{yy}=\bra{\alpha_{yy}}/\alpha_0$ (dotted line), and
$\tilde{\alpha}_{\rm fit}=\alpha_{\rm  fit}/\alpha_0$ (thick gray line,
but only shown in the second panel).
The two open symbols in the top panel indicate that the values
of $\alpha_{xx}/\alpha_0$ are negative.
}\label{palp_Rm26_64}\end{figure}

The resulting values of $\alpha$ are shown in \Fig{palp_Rm26_64}.
For strong imposed magnetic fields, $\Rm\BB_0^2/\Beq^2>1$,
the resulting dependence of $\alpha$ on $B_0$ obeys the standard
catastrophic quenching formula for the case of a uniform magnetic field
\citep{VC92},
\EQ
\alpha_{\rm fit}={\alpha_0\over1+\tilde\Rm\meanBB^2/B_{\rm eq}^2}\quad
\mbox{(for $\meanBB=\BB_0=\const$ only)},
\label{fit}
\EN
where $\alpha_0=-\onethird\urms$ is the relevant kinematic reference
value for fully helical turbulence with negative helicity and $\Rm>1$
\citep{Sur_etal08}.
We treat $\tilde\Rm$ as an empirical fit parameter that
is proportional to $\Rm$ and find that $\tilde\Rm/\Rm\approx0.4$
gives a reasonably good fit; see the dash-dotted line in \Fig{palp_Rm26_64}.
The existence of such an empirical factor might be related to fact
that the relevant quantity could be the width of the magnetic inertial
range, and that this is not precisely equal to $\Rm$.
For $\Rm\BB_0^2/\Beq^2>1$, a similar result is also reproduced using
the test-field method, although $\alpha_{xx}$ is typically somewhat
larger than $\alpha_{\rm imp}$.

For weak imposed magnetic fields, $\Rm\BB_0^2/\Beq^2<1$, apparent
discrepancies are found between the imposed-field method and the
test-field method.
In fact, in the graphical representation in \Fig{palp_Rm26_64}
the results can be subdivided into four different branches that we
refer to as branches X, Y, Z, and YZ.
These names have to do with the orientation of a dynamo-generated
magnetic field.
These dynamo-generated magnetic fields take the form of Beltrami
fields that vary in the $x$, $y$, and $z$ directions for branches
X, Y, and Z, while for branch YZ the field varies both in the
$y$ and $z$ directions.
Earlier work without imposed fields has shown that branch YZ can
be accessed during intermediate times during the saturation of the
dynamo, but it is not one of the ultimate stable branches X, Y, or Z.

Branches Y and Z show the sudden onset of suppression of
$\alpha_{\rm imp}$ for {\it weak} magnetic fields.
This has to do with the fact that for weak imposed magnetic fields
a dynamo-generated field of Beltrami type is being generated.
Such fields quench the $\alpha$ effect, even though they do not
contribute to the volume-averaged mean field.
On branch YZ the $\alpha$ effect is only weakly suppressed, while
on branch X the imposed-field $\alpha_{\rm imp}$ increases with
decreasing values of $B_0$.

The test-field method reveals that on branches X, Y, and YZ the
$\alpha_{yy}$ component is nearly independent of $B_0$, and
always larger than the $\alpha_{xx}$ component.
However, on branch Z and for $\Rm\BB_0^2/\Beq^2<1$ we find that
$\alpha_{xx}=\alpha_{yy}$ and only weakly suppressed.

\begin{figure}\begin{center}
\includegraphics[width=\columnwidth]{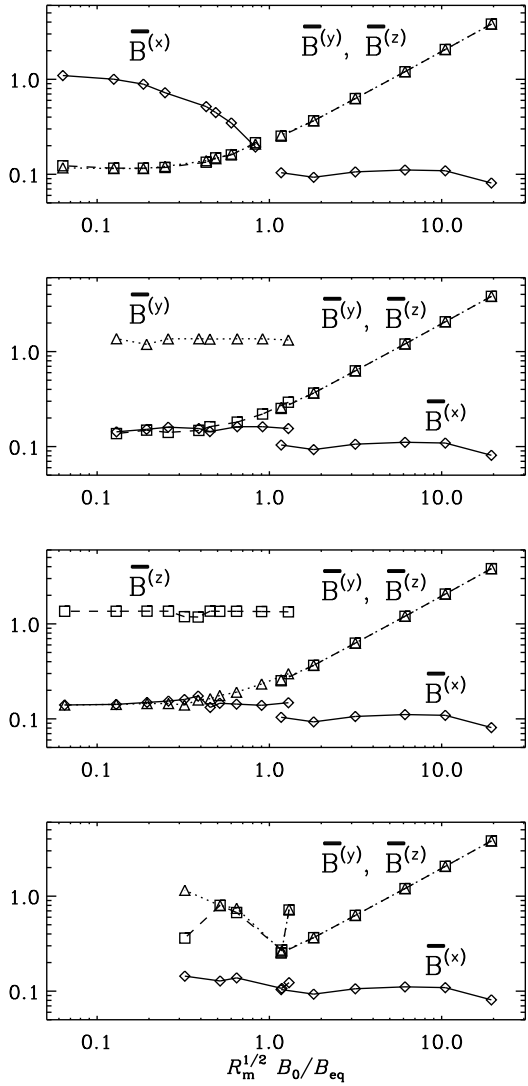}
\end{center}\caption[]{
Root-mean-square values of the mean magnetic fields as functions of
the imposed field for turbulence with $\Rm=26$ for the X, Y, Z, and YZ
branches in the same order as in \Fig{palp_Rm26_64}.
Diamonds, triangles, and squares denote
$\meanBB^{(x)}$, $\meanBB^{(y)}$, and $\meanBB^{(z)}$, respectively.
}\label{pbmxyz_Rm26_64}\end{figure}

\begin{figure}\begin{center}
\includegraphics[width=\columnwidth]{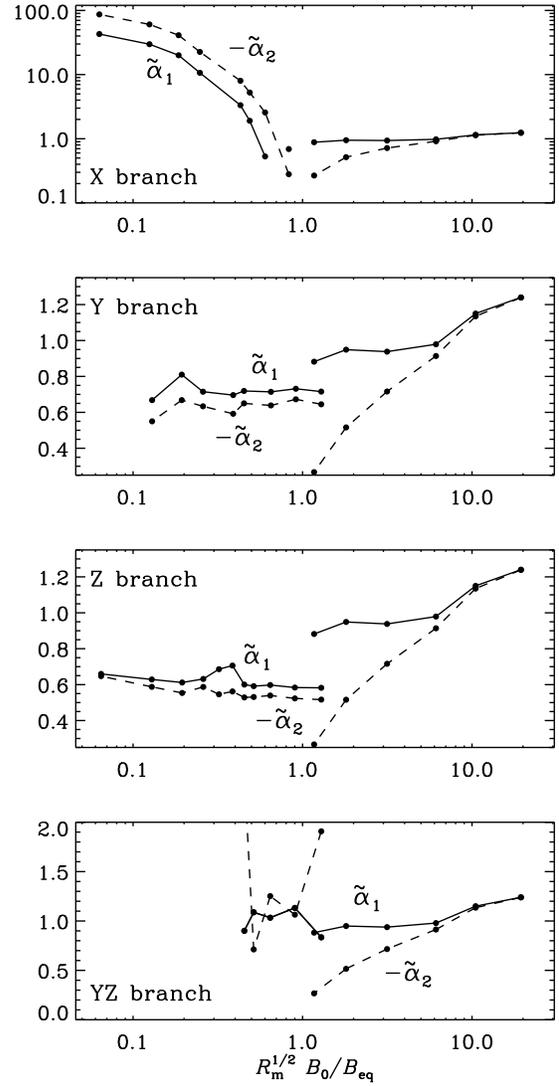}
\end{center}\caption[]{
Dependence of $\alpha_1$ and $\alpha_2$ on $B_0$ for the X, Y, Z, and
YZ branches in the same order as in \Fig{palp_Rm26_64}.
}\label{palp12_Rm26_64}\end{figure}

A comment regarding the discontinuities in \Fig{palp_Rm26_64}
near $\Rm\BB_0^2/\Beq^2=1$ is here in order.
The systems considered here are in saturated states.
To the left of the discontinuities
the system has a saturated meso-scale dynamo, while to the right
there is none.
Intermediate states are simply not possible.
Hence, the discontinuities are caused by the effects of the
meso-scale magnetic fields on $\urms$ and thus on $\Rm$.

\subsection{Relation to $\alpha_1$ and $\alpha_2$}

In the following we will try to interpret the results presented above
in terms of \Eq{alpij} and determine $\alpha_1$ and $\alpha_2$ for the
different branches.
For small values of $B_0$, a magnetic field
with $k=k_1$ and hence a finite planar average can develop.
Compared with the large-scale field $B_0$, we refer to this
dynamo-generated field as meso-scale magnetic field.
As demonstrated in \cite{B01}, three types of such mean fields are possible
in the final saturated state.
These fields correspond to Beltrami fields of the form
\EQ
{\meanBB^{(x)}\over B_1}=\pmatrix{0\cr c_x\cr s_x},\quad
{\meanBB^{(y)}\over B_1}=\pmatrix{s_y\cr0\cr c_y},\quad
{\meanBB^{(z)}\over B_1}=\pmatrix{c_z\cr s_z\cr0},\quad
\EN
where $c_\xi=\cos(k_1\xi+\phi)$ and
$s_\xi=\sin(k_1\xi+\phi)$ denote cosine and sine functions
as functions of $\xi=x$, $y$, or $z$, with an arbitrary phase shift
$\phi$.\footnote{
Unlike the case considered by \cite{BRRS08}, here 
the test field has $k=0$, and there is no relative phase
to be considered.}
The precise value of $B_1$ emerges as a result of the simulation,
but based on simulations in a periodic domain \citep{B01} we know that
$B_1/\Beq$ should be about $(\kf/k_1)^{1/2}$ times the equipartition value.
This is also confirmed by the present calculations.

Let us now discuss separately the different branches.
As can be seen from \Fig{pbmxyz_Rm26_64}, the weak-field regime
is characterized by the presence of meso-scale magnetic fields
that vary either in the $x$ direction (the X branch),
the $y$ direction (Y branch), the $z$ direction (Z branch),
or in both the $y$ and $z$ directions (YZ branch).

In order to get some idea about the values $\alpha_1$ and $\alpha_2$
on the various branches, we consider two limiting cases.
For strong imposed fields, $\beta\to\infty$, the results lie formally on the
YZ branch branch, because such a field has only very little variation
in the $x$ direction.
However, $\bra{\hatB_i\hatB_j}$ will be dominated only by the uniform
field in the $x$ direction, so
we have $\epsilon_x=1$ and $\epsilon_y=0$; see \Sec{testfield}.
This means that $\tilde\alpha_{\rm imp}=\tilde\alpha_{xx}=\tilde\alpha_1+\tilde\alpha_2$
and $\tilde\alpha_{yy}=\tilde\alpha_1$, so we can calculate
\EQ
\tilde\alpha_1=\tilde\alpha_{yy},\quad
\tilde\alpha_2=\tilde\alpha_{\rm imp}-\tilde\alpha_{yy},
\EN
where a tilde indicates normalization by $\alpha_0$.
For weak imposed fields, $\beta\to0$, we can calculate
$\tilde\alpha_1$ and $\tilde\alpha_2$ 
on the X branch by using using the relations
\EQA
\tilde\alpha_{xx}&\!=&\!\tilde\alpha_1,\label{alpxx}\\
\tilde\alpha_{yy}&\!=&\!\tilde\alpha_1+\half\tilde\alpha_2,\label{alpyy}\\
\tilde\alpha_{\rm imp}&\!=&\!\tilde\alpha_1+\tilde\alpha_2\label{alpimp}.
\ENA
However, on the X branch $\tilde\alpha_{xx}$ is ill-determined,
as seen in \Fig{palp_Rm26_64} and discussed in \Sec{Effectiveness} below.
Therefore we use only \Eqs{alpyy}{alpimp} to calculate
\EQ
\tilde\alpha_1=2\tilde\alpha_{yy}-\tilde\alpha_{\rm imp},\quad
\tilde\alpha_2=2\tilde\alpha_{\rm imp}-2\tilde\alpha_{yy}.
\EN
For the Y, Z, and YZ branches, on the other hand, these relations
have to be substituted by
\EQ
\tilde\alpha_1=2\tilde\alpha_{xx}-\tilde\alpha_{\rm imp},\quad
\tilde\alpha_2=2\tilde\alpha_{\rm imp}-2\tilde\alpha_{xx}.
\label{alp12onYZbranches}
\EN
The resulting values of $\tilde\alpha_1$ and $\tilde\alpha_2$ are plotted in
\Fig{palp12_Rm26_64} for each of the four branches.
On the Y branch one can, as a test, also use the independent relation
$\tilde\alpha_1=\tilde\alpha_{yy}$.
The resulting values are about 50\% larger than the values shown
in \Fig{palp12_Rm26_64}, suggesting that there could be additional
contributions in the simplified relation $\tilde\alpha_{yy}=\tilde\alpha_1$.
On the Z branch, of course, $\tilde\alpha_{xx}=\tilde\alpha_{yy}$,
so here too we have to use the equations \eq{alp12onYZbranches}.

In all cases we find that $\tilde\alpha$ is quenched by $\tilde\alpha_1$ and $\tilde\alpha_2$
having opposite signs and their moduli approaching each other.
This is particularly clear in the case of strong fields where
$\tilde\alpha_1$ and $-\tilde\alpha_2$ become indistinguishable,
while each of them is still increasing.
We note that the turbulence itself is not strongly affected \citep{BS05b}.
On the Y and Z branches both $\tilde\alpha_1$ and $\tilde\alpha_2$ are of order unity,
but on the X branch they can reach rather large values when the imposed
field is weak.
The behavior on the YZ branch is somewhat unsystematic, suggesting that
this branch is really just the result of a long-term transient, as was
already found in the absence of an imposed field \citep{B01}.
However, we decided not to discard this branch, because it is likely
that transient solutions on this branch may become even more long-lived
as the magnetic Reynolds number is increased further.

\subsection{Enhancement of $\alpha_{\rm imp}$ in the field-aligned case}
\label{Enhancement}

The suppression of $\alpha=\alpha_1+\alpha_2$ by the magnetic field is
not surprising.
What is unexpected, however, is the dramatic enhancement of both $\alpha_1$
and $-\alpha_2$ for weak imposed fields and equipartition-strength
meso-scale fields that vary in the $x$ direction
(the field-aligned case or X branch).
In this case the interactions of the current density associated with
the Beltrami field and the imposed field generate a force varying along $x$,
perpendicular to the components of the meso-scale Beltrami field.
This generates a meso-scale velocity that in turn damps the Beltrami field,
resulting in the slower rise in $\meanBB^{(x)}$ as $B_0/\Beq$ is decreased.
Further, the cross-product of the meso-scale velocity field with the
Beltrami field generates a large-scale electromotive force in the $x$ direction.
This is seen both in $\alpha_{\rm{imp}}$ and in $\alpha_{xx}$.
A rough estimate of this electromotive force can be obtained by
considering the fields
\EQ
\BB_0=\pmatrix{B_0 \cr 0\cr 0}, \quad
\BB_1=B_1 \pmatrix{0 \cr \cos kx \cr \sin kx}, \quad
\EN
so that $\mu_0\JJ_1=-k\BB_1$, where subscripts 1 denote meso-scale fields.
The meso-scale current density and the imposed field will
generate a meso-scale Lorentz force which will drive a meso-scale velocity
field $\UU_1$.
We estimate $\UU_1$ by balancing
\EQ
\JJ_1 \times \BB_0 /\rho + \nut \nabla^2 \UU_1\approx0,
\EN
where $\nu_t$ is the turbulent viscosity.
We therefore expect that $\UU_1$ will saturate for
\EQ
\UU_1=\frac{B_0 B_1/\rho\mu_0}{\nut k}
\pmatrix{0 \cr \sin kx \cr  -\cos kx }.
\EN
This velocity field will generate an $\EMF_0$ parallel to $\BB_0$ in conjunction with $\BB_1$  
\EQ
\EMF_0 \equiv \bra{\UU_1\times\BB_1} = \alpha_{\rm meso}\BB_0,
\EN
with $\alpha_{\rm meso}=B_1^2/(\rho\mu_0\nut k)$.
We then expect the total $\alpha_{\rm{imp}}$ to be
\EQ
\alpha_{\rm{imp}}=\alpha+\frac{B_1^2/\rho\mu_0}{\nut k}.
\EN
Normalizing by $\alpha_0=-\urms/3$ and assuming $\nut\approx\urms/3\kf$
we find for small imposed field and a meso-scale dynamo that varies along $x$:
\EQ
\frac{\alpha_{\rm{imp}}}{\alpha_0}\approx
1+9\,{\kf\over k_1}\left(\frac{B_1}{\Beq}\right)^2.
\EN
Given that $\kf/k_1=3$ and noting that $B_1/\Beq$ reaches values
up to 1.2, we find that $\alpha_{\rm imp}/\alpha_0\approx40$, which
is still somewhat below the actual value of 53, see the top panel of
\Fig{palp_Rm26_64}.
The remaining discrepancy may be explicable by recalling that the
actual value of $\nut$ may well be reduced due to the presence of
an equipartition-strength magnetic field.

\subsection{Comment on wavenumber dependence}

In previous work on the test-field method we used test fields
with wavenumbers different from zero.
It turned out that in the kinematic regime, $\alpha$ is proportional
to $1/[1+a(k/\kf)^2]$, where $a=0.5, ..., 1$ \citep{BRS08,MKTB09}.
It was shown that the variation of $\alpha$ with $k$ represents
nonlocality in space.
In order to get some idea about the dependence of $\alpha_{xx}$ and
$\alpha_{yy}$ on $k$ in the present case we compare in \Tab{Tabk}
the results for $k=k_0$ with those for $k=0$.
It turns out that both values decrease by 30\% on the X branch, and
increase by less than 10\% on the Z branch.

\begin{table}\caption{
Examples of the dependence of $\tilde\alpha_{xx}$ and $\tilde\alpha_{yy}$
on the wavenumber $k$ of the test field.
Note that the field strength is different in both cases.
}\vspace{12pt}\centerline{\begin{tabular}{ccccc}
Branch & $k/k_0$ & $\tilde\alpha_{xx}$ & $\tilde\alpha_{yy}$ &
$\Rm^{1/2}B_0/\Beq$\\
\hline
X & 0 & $ 0.72\pm0.14$ & $0.51\pm0.16$ & 0.06\\
  & 1 & $ 0.61\pm0.02$ & $0.37\pm0.01$ & 0.06\\
Z & 0 & $ 0.34\pm0.02$ & $0.32\pm0.02$ & 0.2\\
  & 1 & $ 0.35\pm0.01$ & $0.35\pm0.02$ & 0.2\\
\label{Tabk}\end{tabular}}\end{table}

The $k$ dependence for the Z branch is minor, although one would have
expected a small decrease rather than an increase.
Nevertheless, within error bars, this result is
possibly still compatible with the dependence in the kinematic case.
For the X branch the error bars for $k=0$ are larger.
This is because of the strong interaction between the imposed uniform
field and a Beltrami field varying along the same direction, as discussed
in \Sec{Enhancement}.
It is therefore not clear whether the $k$ dependence is here significant and
how to interpret it.

\section{Resetting the fluctuations}
\label{Resetting}

\subsection{Effectiveness of resetting the fields}
\label{Effectiveness}

The evolution equations used both in the imposed-field method and in
the test-field method allow for dynamo action.
This led \cite{OSBR02} and \cite{KKOS06} to the technique of resetting
the resulting magnetic field in regular intervals.
This method is now also routinely used in the test-field approach
\citep{Sur_etal08}, and we have also used it throughout this work.
The lack of resetting the magnetic field may also be the main reason for
the rather low values of $\alpha$ found in the recent work of \citep{HP09};
see the corresponding discussion in \cite{Ketal09b}.

\begin{figure}\begin{center}
\includegraphics[width=\columnwidth]{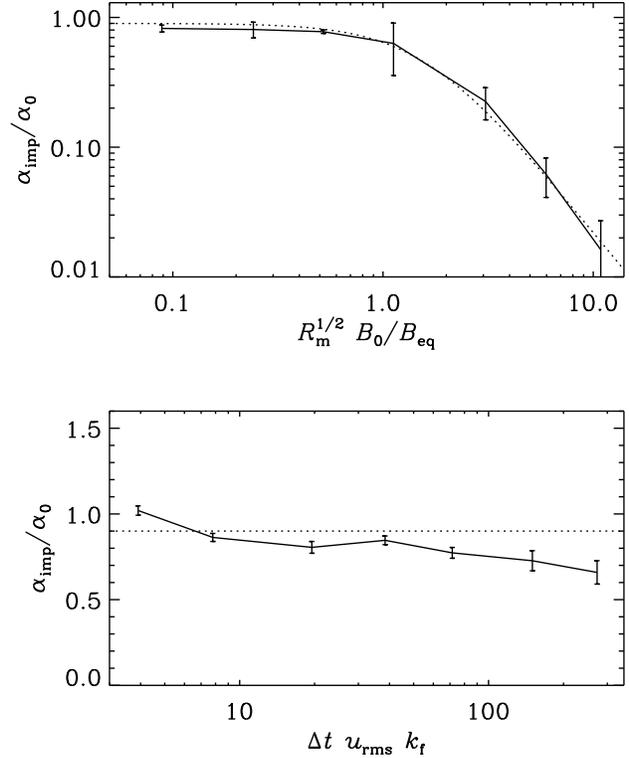}
\end{center}\caption[]{
Dependence of $\alpha_{\rm imp}$ (solid lines)
and $\alpha_{\rm fit}$ (dotted lines) on the imposed field strength
with fixed reset time $\Delta t\urms\kf=50, ..., 70$ (upper panel)
and the dependence of $\alpha_{\rm imp}$ on the reset time
for $\Rm^{1/2}B_0/\Beq=0.1$ (lower panel).
In all cases we have $\Rm\approx30$.
}\label{palp_reset}\end{figure}

In this section we employ the method of resetting $\BB$ to obtain
better estimates for $\alpha$ for weak imposed fields, and to compare
this with results from the test-field method.
The result is shown in \Fig{palp_reset} where we show the dependence
of $\alpha_{\rm imp}$ on $B_0$ and on the reset interval $\Delta t$.
We note that, in units of the turnover time, the reset interval
$\Delta t\urms\kf$ has a weak dependence both on $B_0$ and $\Delta t$,
because small values of $B_0$ and $\Delta t$ quench $\urms$ only weakly.
The resetting technique has eliminated the branching for weak fields.
For weak fields we find that the value of $\alpha_{\rm imp}$ is slightly below
$\alpha_0$, but this is partly because for finite scale separation there is
an additional factor $(1+\kf^2/k_1^2)^{-1}\approx0.9$ \citep{BRS08}.
The actual value of $\alpha_{\rm imp}$ is somewhat smaller still,
which may be ascribed to other systematic effects.

\begin{figure}\begin{center}
\includegraphics[width=\columnwidth]{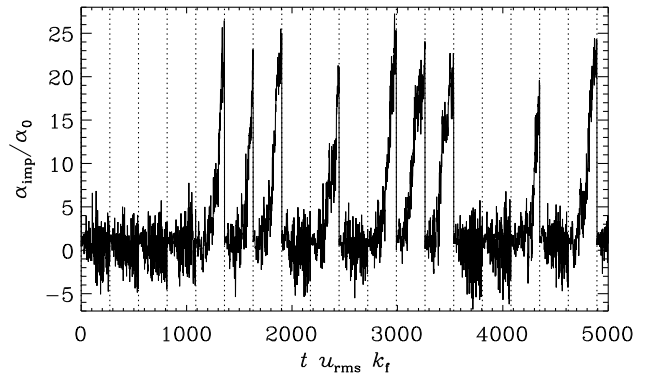}
\end{center}\caption[]{
Time series of $\alpha_{\rm imp}$ for $\Delta t\urms\kf=270$
with $\Rm^{1/2}B_0/\Beq=0.1$.
The reset intervals are indicated by dotted vertical lines.
In all cases we have $\Rm\approx30$.
}\label{palp_impt}\end{figure}

It turns out that over a wide range of reset intervals
the resulting values of $\alpha_{\rm imp}$ are not
dependent in a systematic way on the reset interval
\citep[see also][]{MKTB09}, although it is
clear that the error bars increase for larger values of $\Delta t$.
The same is true for the values of $\alpha_{xx}$ and $\alpha_{yy}$
obtained using the test-field method, except for the case of weak fields
on the X branch where the values of $\alpha_{xx}$ are ill-determined; see
\Tab{Tab}, where we compare the values of $\alpha_{xx}$ and $\alpha_{yy}$
for two different reset times in the case where $\alpha_{xx}$ is found
to change sign ($\Rm^{1/2}\BB_0/\Beq\approx0.2$).
The increasing fluctuations for longer reset intervals occur as the
system exits the kinematic regime.
It might therefore be possible to find indicators of when the
kinematic regime has been exited and resetting becomes necessary.
However, we have not pursued this further in this work.

For even larger values of $\Delta t$ there is enough time for the
meso-scale magnetic field to develop.
An example is shown in \Fig{palp_impt} where 18 intervals of length
$\Delta t\urms\kf=270$ are shown.
For half of these intervals the wavevector of the Beltrami field begins
to develop in the $x$ direction, so $\alpha_{\rm imp}$ is heading
toward the X branch.
In the other half of these cases the magnetic field is weak and
$\alpha_{\rm imp}$ lies on one of the other branches.
None of these cases reproduce the correct kinematic value of $\alpha$,
because we are not really considering a kinematic problem in this case.
This underlines the importance of choosing reset intervals that are not
too long.

Our results support the hypothesis that the precise value of the reset
time interval is not critical except for the field-aligned case where
the diagonal components of the $\alpha_{ij}$ tensor are large and quite
uncertain, as indicated also by the large error bars.
The sign-change found for $\alpha_{xx}$ at low or intermediate field
strengths might therefore not be real.

\begin{table}\caption{
Comparison of the results for $\tilde\alpha_{xx}$ and $\tilde\alpha_{yy}$
for two different reset times $\Delta t$
for the examples of the X and Z branches with $\Rm^{1/2}B_0/\Beq=0.2$.
The reset time is normalized by the inverse turnover time $(\urms\kf)^{-1}$.
}\vspace{12pt}\centerline{\begin{tabular}{cccc}
Branch & $\Delta t\,\urms\kf$ & $\tilde\alpha_{xx}$ & $\tilde\alpha_{yy}$ \\
\hline
X & 25 & $-0.08\pm0.13$ & $0.54\pm0.02$ \\
  & 50 & $-0.98\pm0.09$ & $0.70\pm0.04$ \\
Z & 25 & $ 0.34\pm0.02$ & $0.32\pm0.02$ \\
  & 50 & $ 0.32\pm0.01$ & $0.33\pm0.03$ \\
\label{Tab}\end{tabular}}\end{table}

\subsection{Time averaging in the test-field method}

We have already demonstrated that the length of the reset interval
is not critical for the value of $\alpha$, but longer reset times
tend to lead to larger errors.
In the present section we demonstrate this for the test-field method
using the idealized case where the turbulent flow velocity is
replaced by simple stationary flow given by the equation
\EQ
\UU=\kf\varphi\zzz+\nab\times(\varphi\zzz),
\EN
with
\EQ
\varphi=\varphi(x,y)=u_0\cos k_0x\cos k_0y,
\EN
which is known as the Roberts flow.

When the magnetic Reynolds number exceeds a certain critical value of
around 60, some kind of dynamo action of $\bb^q$ commences.
This type of dynamo is often referred to as small-scale dynamo action
\citep{Sur_etal08,BRRS08,CH09}, but this name may not always be accurate.
In the case of the Roberts flow there would be no such dynamo action
if the wavenumber of the test field is zero, $k=0$, as assumed here.
However, for $k=k_0$, for example, dynamo action for the test-field
equation is possible.
The test fields are therefore chosen to be
\EQ
{\meanBB^1\over{\cal B}}=\pmatrix{\cos kz\cr 0\cr0},\quad
{\meanBB^2\over{\cal B}}=\pmatrix{\sin kz\cr 0\cr0},\quad
\EN
\EQ
{\meanBB^3\over{\cal B}}=\pmatrix{0\cr\cos kz\cr0},\quad
{\meanBB^4\over{\cal B}}=\pmatrix{0\cr\sin kz\cr0},
\EN
see \cite{Sur_etal08}.
Since now the mean fields are also functions of $z$, the term
$\overline{\uu\times\bb^q}$ cannot be omitted in \Eq{nonSOCA}.

\begin{figure}\begin{center}
\includegraphics[width=\columnwidth]{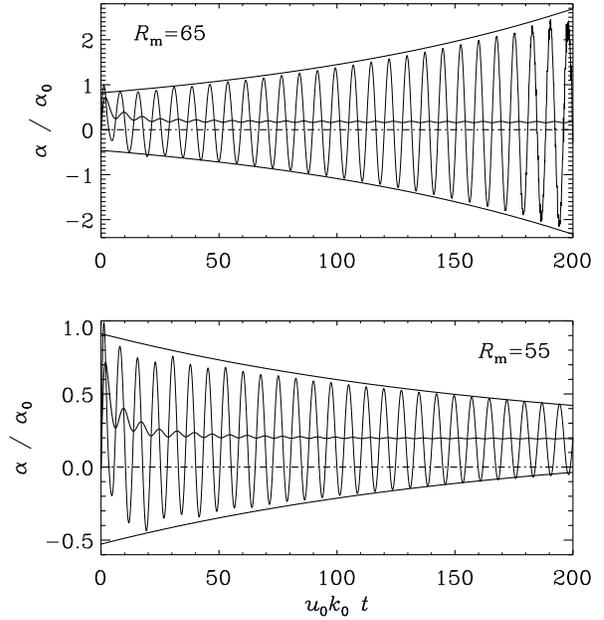}
\end{center}\caption[]{
Plot of the instantaneous $\alpha$ for $\Rm=65$ (upper panel) and
$\Rm=65$ (lower panel).
In both cases running means are overplotted and converge to nearly
the same value of about $-0.096$ in the upper panel and $-0.090$ in the
lower one.
The envelope functions are well described by exponentials and are
also overplotted.
Note however the different scales on the ordinate of both panels.
The dash-dotted line shows the zero level.
}\label{proberts}\end{figure}

As stressed by \cite{BRS08}, in the expression for the electromotive
force there is in general also a contribution $\meanEMF_0$ that is
independent of the mean field.
Given that test fields $\meanBB^q$ are independent of time, we have
\EQ
\meanEMF^q(z,t)=\meanEMF_0^q(z,t)+\alpha(z)\meanBB^q(z)-\etat(z)\mu_0\meanJJ^q(z),
\EN
where overbars denote $xy$ averages (not volume averages), so there is
also a term $\etat\mu_0\meanJJ^q$, where $\etat$ is the turbulent magnetic
diffusivity.
We have assumed that $\alpha$ and $\etat$ are independent of time, and
in this case they are also independent of $z$.
The $\meanEMF_0^q(z,t)$ term can be eliminated by averaging over time,
i.e.\ $\bra{\meanEMF_0^q}=\nullvector$, so
\EQ
\bra{\meanEMF^q}=\alpha\meanBB^q-\etat\mu_0\meanJJ^q.
\EN
In \Fig{proberts} we show the evolution of $\alpha$ for the Roberts flow
with $\Rm=65$ and 55.
In the case with $\Rm=65$ there are exponentially growing oscillations
corresponding to a wave traveling in the $z$ direction.
In general such fields can be a superposition of waves traveling in the
positive and negative $z$ directions.
It is seen quite clearly that the running time average is stable
and well defined.
The results for $\Rm=65$ and 55 are close together ($\alpha/\alpha_0=0.096$
and 0.090, respectively), suggesting continuity across the point where
dynamo action sets in.
This supports the notion that averaging over time is a meaningful procedure.

\section{Conclusions}

The present simulations have shown that the imposed-field method leads to
a number of interesting and unexpected results.
For imposed fields exceeding the value $\Rm^{-1/2}\Beq$ one recovers the
catastrophic quenching formula of \cite{VC92}; see \Eq{fit}.
We emphasize once more, however, that this formula is only valid for
completely uniform large-scale fields in a triply-periodic domain.
This is clearly artificial, but it provides an important benchmark.

A number of surprising results have been found for weaker fields of less
than $\Rm^{-1/2}\Beq$.
In virtually none of those cases does the imposed-field method recover
the kinematic value of $\alpha$.
Instead, $\alpha_{\rm imp}$ can attain strongly suppressed values, but it
can actually also attain strongly enhanced values.
This is caused by the unavoidable emergence of meso-scale dynamo action.
In principle, such meso-scale dynamo action could have been suppressed
by restricting oneself to scale-separation ratios, $\kf/k_1$, of less
than 2 or so.
This was done, for example, in some of the runs of \cite{BS05b}.
In the present case of a triply-periodic box, four different magnetic
field configurations can emerge.
The first three correspond to Beltrami fields, where the wavevector
points in one of the three coordinate directions.
The fourth possibility is also a Beltrami field, but one
that varies diagonally in a direction
perpendicular to the direction of the imposed field.
The latter was found to be unstable in the absence of an imposed field,
but they can be long-lived in the present case of an imposed field.

In this paper, we have used the term meso-scale fields to refer to the
Beltrami fields naturally generated by the helicity-driven dynamo in our system.
A more general definition of meso-scale fields would encompass all fields
that break isotropy, average to zero, and yet do not time-average to zero.
In the absence of such fields, mean-field theory can be applied in a
straightforward manner.
This is indeed the case that one is normally interested in.
However, when such meso-scale fields exist, they must be understood for
determining turbulent transport coefficients, because those coefficients
apply then to the particular case of saturated meso-scale fields.

The results obtained with the imposed-field method reflect correctly
the circumstances in the nonlinear case where the $\alpha$ effect is
suppressed by dynamo-generated meso-scale magnetic fields whose scale
is smaller than that of the imposed field, but comparable to the scale
of the domain.
Especially in the case of closed or periodic domains the resulting
$\alpha$ is catastrophically quenched, which is now well understood
\citep{FB02,BB02}.
This effect is particularly strong in the case where one considers volume
averages, and thus ignores the effects of turbulent magnetic diffusion.
With magnetic diffusion included, both $\alpha$ and $\etat$ have only a
mild dependence on $\Rm$ \citep{BRRS08}.
However, astrophysical dynamos are expected to operate in a regime
where magnetic helicity fluxes alleviate catastrophic
quenching; see \cite{BS05} for a review.

Determining the nature of the dynamo mechanism is an important part in the
analysis of a successful simulation showing large-scale field generation.
Our present analysis shows that meaningful results for $\alpha$ can be
obtained using either the imposed-field or the test-field methods
provided the departure of the magnetic field from $\BB_0$
is reset to zero to eliminate the effects
of dynamo-generated meso-scale magnetic fields.
Conversely, if such fields are not eliminated, the results can still
be meaningful, as demonstrated here, but they need to be interpreted
correspondingly and bear little relation to the imposed field.
On the other hand, for strong imposed magnetic fields
($\Rm\BB_0^2/\Beq^2>1$), meso-scale magnetic fields tend not to grow,
so the resetting procedure is then neither necessary nor would it make
much of a difference when the test-field method is used.
However, when the imposed-field method is used, the resetting of the
actual field reduces the quenching of $\urms$.
This affects the normalizations of $B_0$ and $\alpha_{ij}$
with $\Beq$ and $\alpha_0$, respectively, because both
are proportional to $\urms$.\footnote{
This explains why $\Delta t\urms\kf$ is $70$ in \Fig{palp_reset}
and $50$ in \Tab{Tab} under otherwise comparable conditions, except that
here only the test-field is reset and not the actual fluctuating one.}

Throughout this paper we have considered relatively moderate values
of $\Rm$, but we computed a large number of different simulations.
In the beginning of this study we started with larger values of $\Rm$
and found that the resulting $\alpha_{\rm imp}$ seemed inconsistent.
In hindsight it is clear what happened: the few cases that we had in
the beginning were all scattered around different branches.
Only later, by performing a large number of simulations at smaller values
of $\Rm$ it became clear that there are indeed different branches.
This highlights the importance of studying not just one or a few models
of large $\Rm$, but rather a larger systematic set of intermediate
cases of moderate $\Rm$ where it is possible to understand in detail
what is going on.
It will be important to continue exploring the regime of larger $\Rm$,
and we hope that the new understanding that emerged from studying
cases of moderate $\Rm$ proves useful in this connection.
According to the results available so far, we can say that for larger
values of $\Rm$ the turbulent transport coefficients are only weakly
affected \citep[see][for $\Rm\leq600$]{BRRS08} for fields of equipartition
strength, or not affected at all \citep[][for $\Rm\leq220$]{Sur_etal08}
if the field is in the kinematic limit.

\section*{Acknowledgments}
We thank the referee for making a number of constructive remarks on the paper.
We acknowledge the use of computing time at the Center for
Parallel Computers at the Royal Institute of Technology in Sweden
and CSC -- IT Center for Science in Espoo, Finland.
This work was supported in part by
the European Research Council under the AstroDyn Research Project 227952 (FDS).
the Academy of Finland grant 121431 (PJK),
and the Swedish Research Council grant 621-2007-4064 (AB).


\vfill\bigskip\noindent\tiny\begin{verbatim}
$Header: /var/cvs/brandenb/tex/fabio/ImposedFieldMethod/paper.tex,v 1.89 2009-11-10 13:45:39 brandenb Exp $
\end{verbatim}

\end{document}